\documentclass[aip,reprint,amsmath,amssymb,floatfix]{revtex4-2}

\usepackage{ifthen,commath,graphicx,dcolumn,bm,hyperref,braket,siunitx,natbib}
\usepackage{xcolor,ulem}

\begin{document}

\newcommand{\up}{|\uparrow\rangle}
\newcommand{\down}{|\downarrow\rangle}
\newcommand{\Yb}{$^{171}$Yb$^+\ $}
\newcommand{\ketu}{$ \left| \uparrow \right\rangle \ $}
\newcommand{\ketd}{$ \left| \downarrow \right\rangle \ $}
\newcommand{\ketU}{$ \left| \Uparrow \right\rangle \ $}
\newcommand{\ketD}{$ \left| \Downarrow \right\rangle \ $}

\title[High stability cryogenic system for quantum computing with compact packaged ion traps]{High stability cryogenic system for quantum computing with compact packaged ion traps}

\author{Robert F. Spivey}
\affiliation{Department of Electrical and Computer Engineering, Duke University, Durham, NC 27708}
\author{Ismail V. Inlek},
\affiliation{Department of Electrical and Computer Engineering, Duke University, Durham, NC 27708}
\affiliation{IonQ, Inc., College Park, MD 20740}
\author{Zhubing Jia}
\affiliation{Department of Physics, Duke University, Durham, NC 27708}
\author{Stephen Crain}
\affiliation{Department of Electrical and Computer Engineering, Duke University, Durham, NC 27708}
\affiliation{IonQ, Inc., College Park, MD 20740}
\author{Ke Sun}
\affiliation{Department of Physics, Duke University, Durham, NC 27708}
\author{Junki Kim},
\affiliation{Department of Electrical and Computer Engineering, Duke University, Durham, NC 27708}
\author{Geert Vrijsen}
\affiliation{Department of Electrical and Computer Engineering, Duke University, Durham, NC 27708}
\author{Chao Fang}
\affiliation{Department of Electrical and Computer Engineering, Duke University, Durham, NC 27708}

\author{Colin Fitzgerald,$^{4}$ Steffen Kross,$^{4}$ Tom Noel}
\affiliation{ColdQuanta, Inc., Boulder, CO 80301}
\author{Jungsang Kim}
\affiliation{Department of Electrical and Computer Engineering, Duke University, Durham, NC 27708}
\affiliation{IonQ, Inc., College Park, MD 20740}

\date{\today}%

\begin{abstract}
	Cryogenic  environments benefit ion trapping experiments by offering lower motional heating rates, collision energies, and an ultra-high vacuum (UHV) environment for maintaining long ion chains for extended  periods  of  time. Mechanical vibrations caused by compressors in closed-cycle cryostats can introduce relative motion between the ion and the wavefronts of lasers used to manipulate the ions. Here,  we  present  a  novel  ion  trapping system where a commercial low-vibration closed-cycle cryostat is used in a custom monolithic enclosure. We measure mechanical vibrations of the sample stage using an optical interferometer, and observe a root-mean-square relative displacement of 2.4 nm and a peak-to-peak displacement of 17 nm between free-space beams and the trapping location. We packaged a surface ion trap in a cryo-package assembly that enables easy handling,  while creating a UHV environment for the ions. The trap cryo-package contains activated carbon getter material for enhanced sorption pumping near the trapping location, and source material for ablation loading. Using \(^{171}\)Yb\(^{+}\) as our ion we estimate the operating pressure of the trap as a function of package temperature using phase transitions of zig-zag ion chains as a probe. We measured the radial mode heating rate of a single ion to be 13 quanta/s on average. The Ramsey coherence measurements yield 330 ms coherence time for counter-propagating Raman carrier transitions using a 355 nm mode-locked pulse laser, demonstrating the high optical stability. 
\end{abstract}
\pacs{}

\maketitle

\section{Introduction}

Ion traps are an experimentally verified platform for the storage and manipulation of high fidelity quantum states\cite{wineland1998experimental, monroe2013scaling,blatt2008entangled}. The atomic ions provide fundamentally stable qubits when adequate hyperfine levels are used \cite{langer2005long, harty2014high, wang2017single}. High-fidelity state preparation and detection \cite{myerson2008high, noek2013high, crain2019high}, as well as logic gates driven by both lasers \cite{harty2014high, mount2015error, blume2017demonstration, ballance2016high, gaebler2016high, wang2020high} and microwave fields are readily available\cite{wineland1998experimental}. The performance of trapped ion qubits arises from the fact that these atoms are well isolated from the environment under ultra-high vacuum (UHV) conditions, and the control signals can be delivered to the qubits in the form of electromagnetic fields. 
These operating conditions require typical trapped ion experiments to have a UHV chamber, and a complex set of laser systems to operate.

Traditionally the experimental setups for atomic physics consist of a complex assembly of tabletop optical components, typically assembled individually by hand. While providing ultimate flexibility in the layout of the experiments, such an approach is subject to drift with temperature changes and often requires frequent adjustments to keep the optical alignment optimized. Transforming an ion trapping experiment into a reliable quantum computer is hampered by such stability and reliability concerns. The setup must be broken down into functional blocks, each block being an integrated module, to dramatically improve the system stability. In this paper, we describe a novel system design approach to creating the UHV environment for the trapped ions, and the stable optical system to deliver the necessary laser light to manipulate the ion qubits. 

The centerpiece of our approach is the design, construction, and testing of a compact UHV assembly for traps operating in cryogenic temperatures~\cite{wilpers2013compact,bandi2015demonstration,schwindt2016highly,salim2011compact,hudek2012compact}. The UHV assembly consists of a copper lid sealed to a 100-pin ceramic pin grid array (CPGA) package containing the trap. We refer to this assembly 
as the `trap cryo-package'. The trap cryo-package is cooled to cryogenic temperatures to achieve the pressure levels needed for stable ion trap operation \cite{antohi2009cryogenic, brown2016co}. The trap cryo-package is designed to minimize the volume of the UHV environment to be maintained. It only contains the components that are absolutely necessary for trapping: the ion trap, the atomic source, and activated carbon getter to achieve UHV at cryogenic temperatures. 

Our setup also features a compact optical layout, where each optical function is designed onto a customized breadboard referred to as an `optical block' where the optical elements are precision mounted to a pre-designed reference point. The breadboard can be temperature stabilized to eliminate optical misalignment due to thermal drifts. This design philosophy can be extended to future atomic physics experiments and ion trap quantum computers. 

\section{Experimental Setup}
\begin{figure}[ht]
        \centering
        \includegraphics[width=\columnwidth]{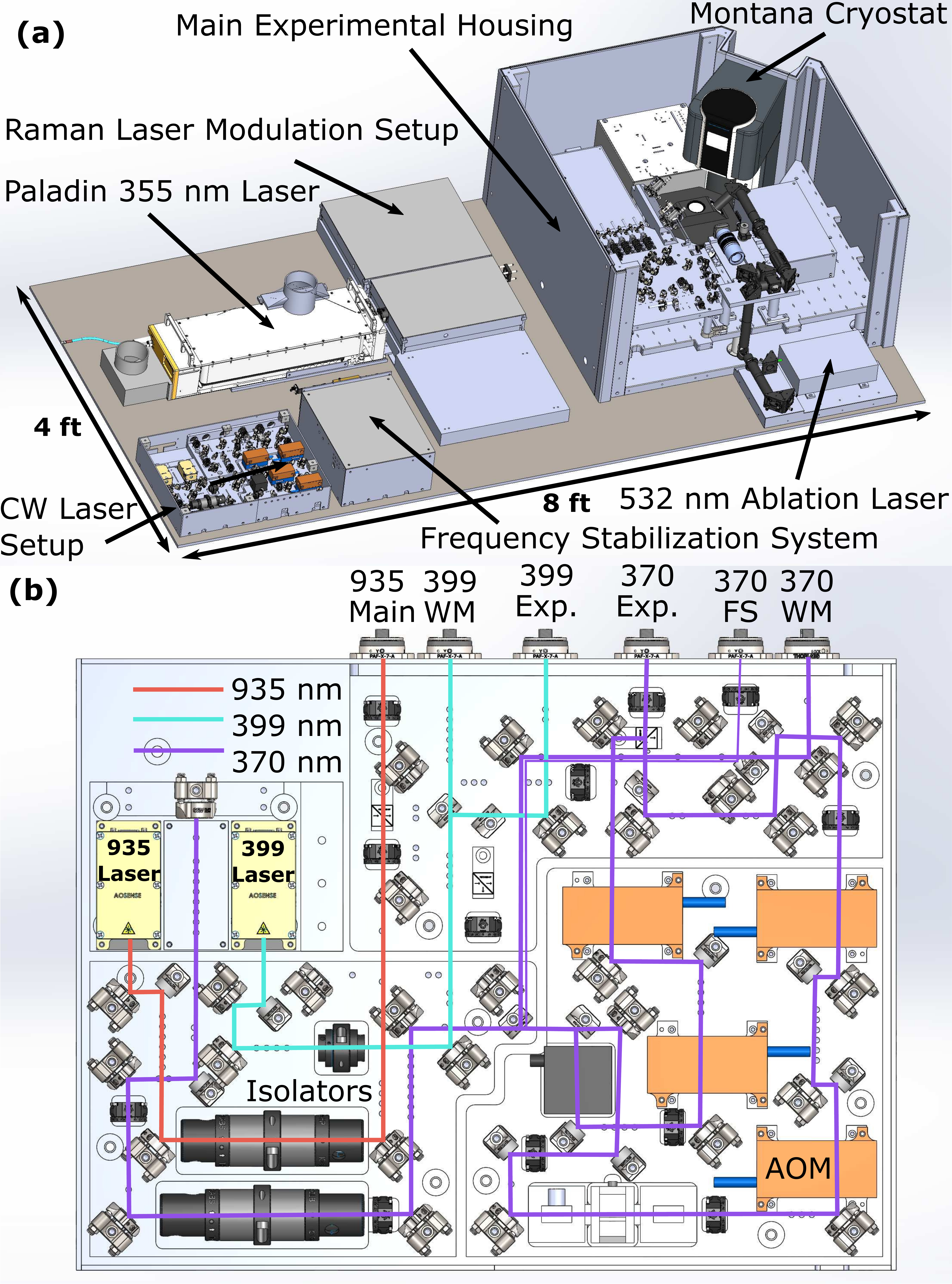}
        \caption{Our optical table setup and system design philosophy. (a) The entire ion trap experiment easily fits onto a single 4 by 8 ft optical table. The setup includes a CW laser launch and modulation plate, a large Raman laser modulation setup which is integrated into a base for the Paladin 355 nm laser, a transfer cavity setup for frequency stabilization, and a large monolithic optical enclosure for the cryostat and main experimental optics. (b) As an example of our optical design strategy, we present the internal design of the continuous wave (CW) laser launch box. It has a larger water cooled base upon which a few optical blocks are mounted. We use 399 nm light for photo-ionization \cite{Vrijsen:19}, 370 nm for \Yb qubit operations, and 935 nm for re-pumping. Availability of fiber products at 935 nm, such as fiber modulators and beamsplitters, simplifies the design for this beam. We use free space splitters, EOM's and AOM's at 370 nm. These custom machined plates allow for compact and highly stable optical assemblies. This CW module measures 48.6 cm (W)$\times$ 36.2 cm (D)$\times$ 9.3 cm (H).}
        \label{fig:WholeTable}
    \end{figure}

	Our overall system consists of four main parts (Figure \ref{fig:WholeTable}(a): a continuous wave (CW) laser setup (shown in more detail in Figure \ref{fig:WholeTable}(b), a laser frequency stabilization system, a 355 nm Raman laser modulation setup, and the main experimental housing. The main experimental housing includes the closed cycle cryostat system and several optical blocks for tasks such as fluorescence collection for ion imaging and state detection, CW beam delivery for non-coherent qubit operations (photo-ionization, Doppler cooling, state initialization and detection), Raman beam delivery for quantum gates, and ablation loading for the ion source \cite{Vrijsen:19}. Laser light between these various modules are routed using optical fibers. This allows for each optical block to be tested independently before they are incorporated into the complete system. 
    	
\subsection{The Cryostat}
    We utilize a closed-cycle cryostat designed for mechanical stability for our operation (Montana Instruments Cryostation s200). This cryostat features low levels of vibration despite the mechanical motion of the Gifford-McMahon cryocooler. Another unique design feature of this cryostat compared to traditional designs is that the sample space is anchored on the tabletop surface rather than hanging from the top\cite{cryoUMD,cryoInns,cryoGV}, which makes swapping of the trap cryo-packages and upgrade of the cryostat sample chamber straightforward. It has the added advantage that optical alignment can be tested before the cryostat is closed up for cooldown. 
    
\begin{figure}[ht]
        \centering
        \includegraphics[width=\columnwidth]{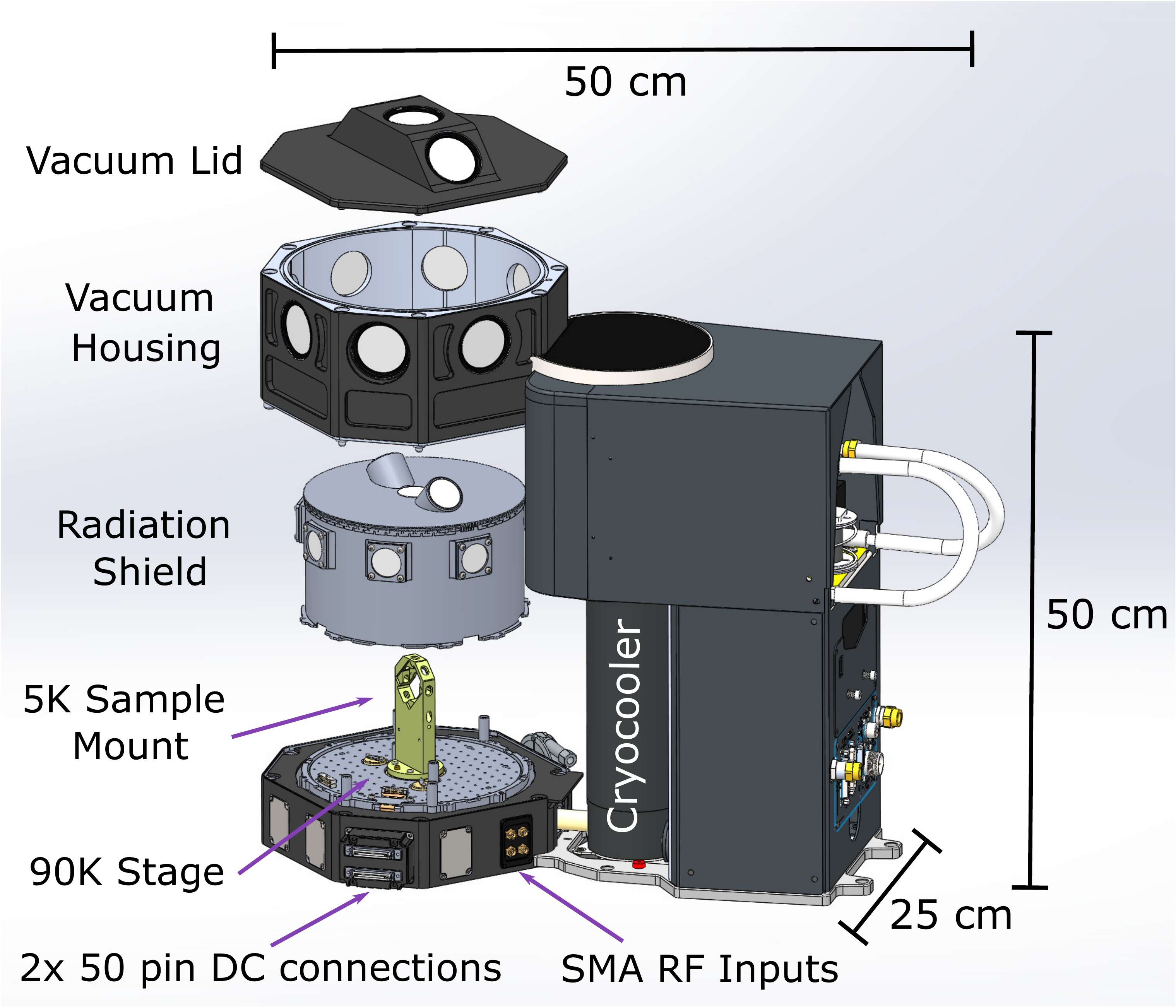}
        \caption{A labelled diagram of the Montana Cryostation system. The vacuum housing and lid with o-ring seals provide the vacuum environment needed for cryogenic operation. The radiation shield is mounted to the 90K stage, and includes all windows necessary for optical access. A gold coated copper sample mount is attached to the 5K cold-finger platform. The trap is mounted within this sample mount such that the trap axis (defined as the x-axis) points along the vertical axis.}
        \label{fig:MontanaCryostat}
    \end{figure}

    A diagram of the cryostat and its internals is shown in Figure~\ref{fig:MontanaCryostat}. The internal sample chamber includes a large flat 90K stage with a grid of tapped holes for mounting various components. In the center there is an exposed 5K base plate, upon which we mount a gold coated copper sample mount that houses and cools our trap cryo-package. A radiation shield with optical viewports through all necessary axes is thermally anchored to the 90K stage. The vacuum housing has two o-ring seals, one making contact with the base of the cryostat, and the other on the top side sealing the lid. Several temperature sensors are installed to monitor the temperature of the system, including the 5K base plate and the sample mount near the trap.   

\subsection{Laser Systems}
	Traditional optomechanical apparatus use posts and clamps to secure optical components, such as mirrors and lenses, to an optical table or breadboard. This strategy offers flexibility in constructing an optical setup but it often results in bulky assemblies. The alignment of these traditional setups drift in response to the changes in the lab environment such as temperature, turbulent air flow, and acoustic noise.
	
	We address this challenge by constructing compact optomechanical modules. Each module serves a self-contained optical function and can be connected to the others via optical fiber. Each module consists of a water cooled base plate with tapped holes and dowel pins upon which optical blocks can be mounted and located. The optical blocks have the tapped holes and dowel pins for aligning the optical components. 

	\subsubsection{Continuous Wave Laser System}
    Figure \ref{fig:WholeTable}(b) shows the continuous wave (CW) lasers needed for trapping, cooling, initializing, and measuring the \(^{171}\)Yb\(^{+}\) hyperfine qubits \cite{olmschenk2007manipulation} used in our experiments. The 399 nm laser (AO Sense. Inc.) is used for resonant photo-excitation of the neutral Yb atoms for loading the trap. For efficient, isotope-selective ablation loading, a second photo-ionization step is required to excite the electron to the continuum. Here, we use the intense beam from the 355 nm Raman laser to achieve this goal. The 370 nm laser is resonant with the main Yb\(^{+}\) S-P transition, and is used for Doppler cooling, optical pumping for qubit initialization, and exciting resonant fluorescence for qubit state detection. The Doppler cooling requires a modulated sideband at 14.7 GHz to pump out the dark state in the ground state manifold, and optical pumping requires a 2.1 GHz frequency shift. These are provided by electro-optic modulators (EOMs) in the beam path. Acousto-optic modulators (AOMs) are used to provide shutter functionality and fine tuning of the cooling and detection frequency. These AOMs can also be used to intensity stabilize our downstream free-space beams which are incident upon the ion. The beam is divided up and fiber-coupled to be sent to either the experiment, the frequency stabilization setup, or to a wavemeter for frequency monitoring. The 935 nm laser is needed to re-pump the atomic population from the meta-stable \(^{2}\)D\(_{3/2}\) states back to the \(^{2}\)S\(_{1/2}\)-\(^{2}\)P\(_{1/2}\) manifold. The optical elements are positioned precisely on the base plate using dowel pins according to a layout generated using computer-aided design (CAD) tools. The base plate is temperature-stabilized with a water chiller, and enclosed to minimize air turbulence along the optical beam paths. This optical setup is stable enough that the single-mode fiber coupling alignments have not required adjustment for over a year of system operation.
    \subsubsection{Frequency Stabilization System}	    
    The frequency of the CW lasers driving atomic transitions (370 nm and 935 nm lasers) must be stabilized to the MHz level or below. This is achieved by taking some of the laser output and locking it to an optical cavity using the Pound-Drever-Hall technique.\cite{drever1983laser} The length of the optical cavity itself is stabilized to an absolute frequency reference, which is a 780 nm laser stabilized against Rb vapor using Doppler-free spectroscopy techniques. The setup is designed and constructed in a similar approach to the CW laser modulation plate shown in Figure \ref{fig:WholeTable}(b).

    \subsubsection{Raman Laser Modulation Setup}
    The coherent operations on the qubits, such as a single and two-qubit gates, are driven by Raman transition using a far-detuned laser at 355 nm.\cite{campbell2010ultrafast} We use a mode-locked, frequency-tripled Nd:YAG laser for this purpose (Coherent Paladin Compact). The output of the laser is split and modulated using multiple AOMs, driven by programmable RF sources to implement the logic gates. In our system, the laser is split into a global beam used to illuminate the entire ion chain from one side, and a pair of individual beams that will be tightly focused onto two selected ions in the chain from the other side, in a counter propagating geometry. The global beam and the pair of individual beams go through their respective modulation paths through the AOMs, and are coupled into a single-mode photonic crystal fiber \cite{colombe2014single}. This modulation setup is called the `upstream' Raman optics. After the fibers, the global beam is shaped into an elongated elliptical beam in the `downstream' optics, and aligned to the ion chain. The individual beams are passed through an addressing system utilizing tilting mirrors fabricated using micro-electromechanical systems (MEMS) technology \cite{knoernschild2009multiplexed, kim2007design,knoernschild2010independent}, where the beam steering functionality provides the ability to address individual ions in the chain with minimal crosstalk \cite{mount2015error,crain2014individual}. The individual beams are also prepared in an elliptical shape to clear the limited numerical aperture available in the surface ion trap chip. 
	    
	\subsection{Main Experimental Enclosure}
	
	The quality of the quantum logic gates imposed on the ion qubits will depend strongly on the stability of the laser beams seen by the qubits that drive the gates. Given that all laser beams are delivered to the ions in the cryostat through optical fibers, the optomechanical structure to accommodate beam path between the fiber and the ion requires care in reducing vibration and drift. The structure is placed in the main experimental enclosure and consists of a monolithic main frame onto which the various `downstream' optical blocks are assembled. All optical blocks and base plates are machined from MIC-6 (Tool and Jig) cast aluminum plates. This material is chosen for its low stress and good flatness specifications. The granular structure of cast aluminum allows for high speed machining while minimizing distortion which can arise when working with rolled aluminum. 
	
	\subsubsection{Optomechanical Setup}
    The entire Montana Instruments cryostat sample chamber and cryocooler module rest atop a 3.5 cm thick base plate which is 0.89 m long on a side. This makes up the foundation of the experimental setup, serving as a mechanical substrate for the ion trap chamber and all downstream optics such that vibrations between them are common-mode. A series of evenly located 1.5 inch thick steel posts are used to suspend a second 1-inch thick optical breadboard at a height for the ion addressing optical block setups to operate. This custom optical breadboard has a grid of 1/4-20 tapped holes with a 2 inch pitch, and each tapped hole has two accompanying 2 mm precision holes for dowel pins. In the space between the base plate and the optical breadboard we have room for routing various cables and vacuum lines to and from the cryostat. 
	    
	\subsubsection{Optical Blocks}
	Each unit of optical functionality is implemented as an optical block, which takes the form of a block of aluminum plate machined on top to locate and fasten optical components in space, similar to the unit shown in Figure \ref{fig:WholeTable}(b). Each optical block is installed onto the optical breadboard using 2 mm dowel pins for precision positioning. By fabricating the blocks from MIC-6 aluminum, good flat-to-flat interfaces between the breadboard and the optical blocks are ensured. Optical components themselves are directly mounted on the aluminum modules with dowel pins for positioning. Some optical modules are elevated with steel posts, where a different height in the enclosure is needed. Custom-designed corner pieces are used to join the large base plate with the optical breadboard, and to attach side and top panels for the housing. Each optical block as well as the entire optomechanical structure can be enclosed with a metallic cover to eliminate air turbulence and reduce acoustic vibration. 
	
\begin{figure}[ht]
        \centering
        \includegraphics[width=\columnwidth]{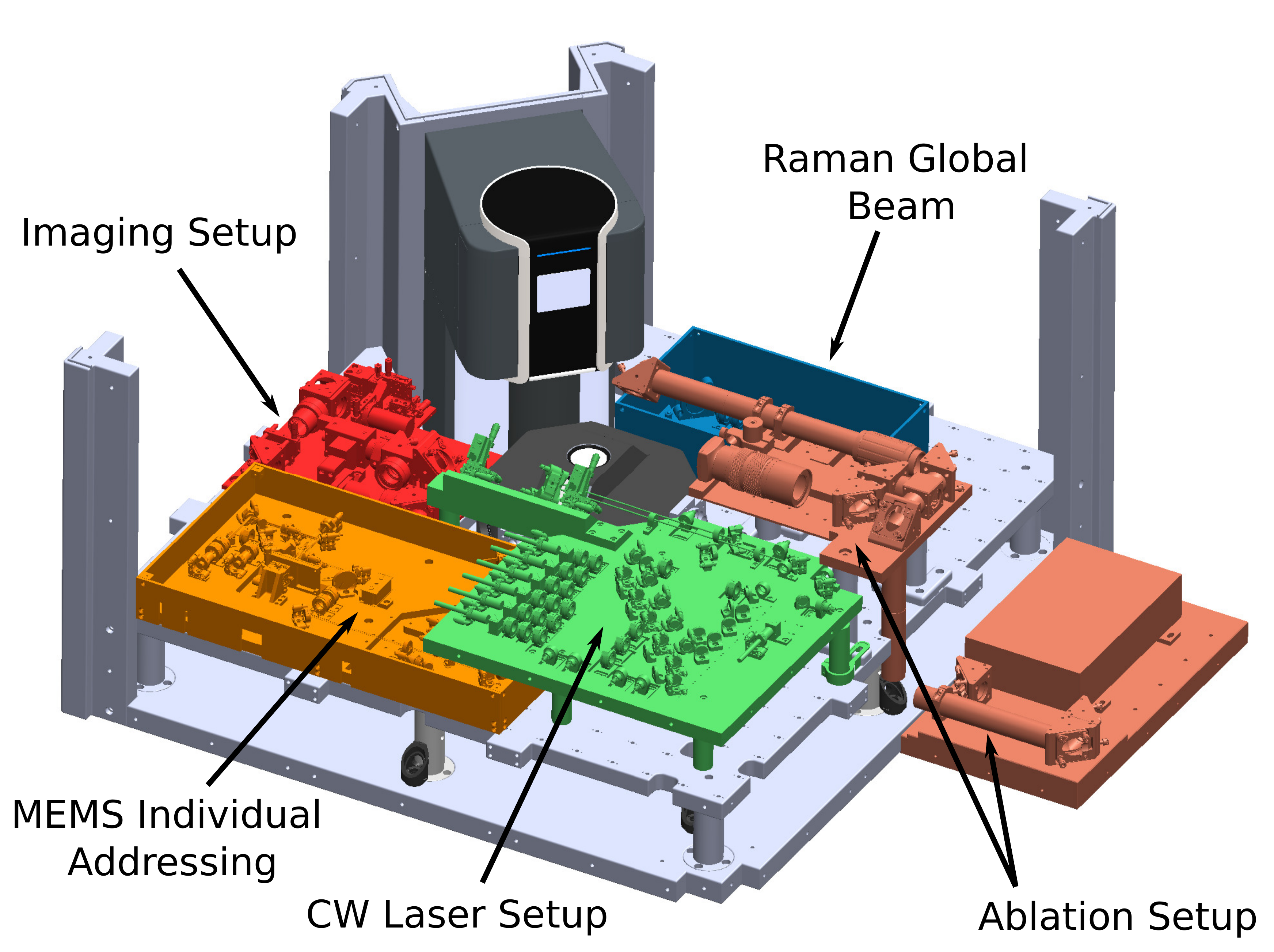}
        \caption{Inside the main experimental enclosure, where the various optical blocks and the Montana cryostation are housed is depicted. (red) is the ion imaging setup, (orange) is the MEMS-based beam steering system for individual Raman addressing, (green) is the CW laser path which includes space for all trapping lasers to be co-propagated, (copper-colored) is the ablation setup which has two parts, where the laser head is attached external to the system and the 532 nm light is brought in via built-in periscope to the upper levels, and (blue) is the global Raman addressing beam. All of these optical blocks are located in space with 2 mm dowel pins and fixed to the optical breadboard with  1/4-20 screws.}
        \label{fig:MontanaHousing}
    \end{figure}

	This modular design strategy combines the stability of monolithic machined pieces with the experimental flexibility needed to re-design and improve optical functions. Each optical block can be replaced and assembled into the optical breadboard with precision alignment with respect to other blocks. Figure \ref{fig:MontanaHousing} shows an image of the optical blocks placed on the optical breadboard, color coded to highlight their optical functions. The optical blocks include: (1) an imaging plate (red) for state fluorescence detection designed to image photons collected from each ion onto a multi-mode fiber in a fiber array, (2) a CW beam delivery plate (green) which co-propagates all trapping lasers and brings them into the ion trap at a downward 45\(^{\circ}\) angle elevated to access the window on the top cover of the cryostat, (3) two Raman optical blocks devoted to the counter-propagating Raman assembly, one for bringing in a global beam (blue) and the other to integrate the MEMs-based beam steering system (orange) for two individual beams, and (4) two optical blocks for the ablation laser (Q-switched Nd:YAG laser, copper-colored), one exterior to the enclosure upon which the laser head is mounted with steering mirrors and a periscope which brings this beam up to the second module which delivers the ablation beam to the atomic source in the trap cryo-package.
	
\subsubsection{Compact Trap Cryo-Package}
\begin{figure}[ht]
        \centering
        \includegraphics[width=\columnwidth]{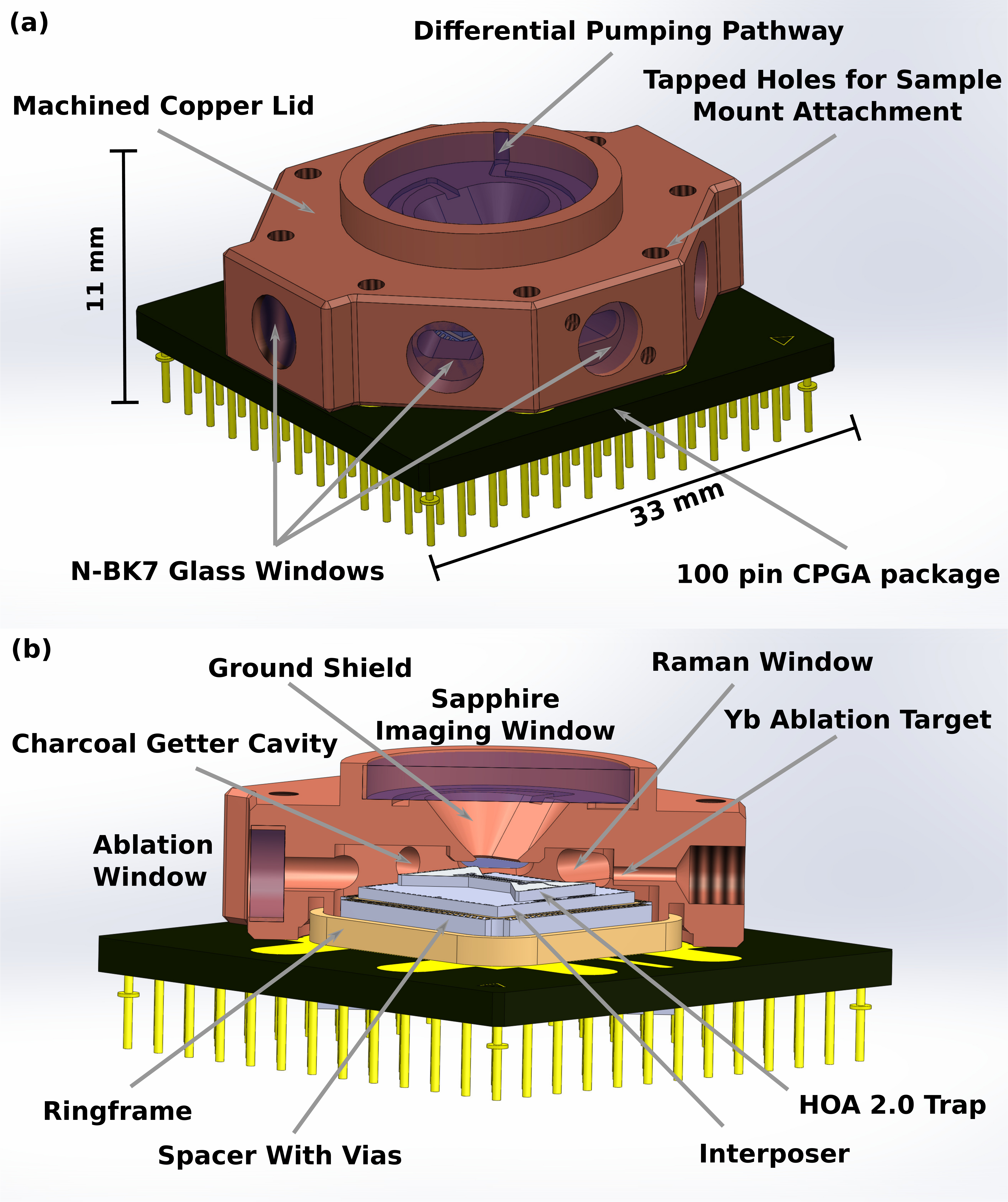}
        \caption{Compact ion trap cryo-package design (a) On the exterior, it features eight tapped holes on the top and a spiral meandering pathway cut into the copper lid for pumping. The imaging window covers and makes the fourth wall of this meandering pathway, which is used for gas evacuation. AR coated N-BK7 side windows are epoxy-sealed to this copper lid. (b) A cross section shows the interior features of this assembly. A sapphire imaging window is epoxy-sealed to the top. The internal features include a cavity for the storage of carbon getter packaged in a copper mesh, and a holder for storage of the Yb ablation target. The ion trap stack is packaged with an interposer for routing wire bonds, and a spacer to raise the top surface of the trap to clear the ring-frame height in the lid.}
        \label{fig:TrapPackage}
    \end{figure}

    The vacuum inside the cryostat is maintained with o-ring seals, as opposed to copper gaskets which are typically used in UHV-chambers. This sealing mechanism limits the pressure to the low $10^{-8}$ Torr range. To achieve UHV vacuum levels in the trapping region, we create a secondary chamber assembly called the `trap cryo-package' to hold the trap, which supports a UHV environment within the cryostat via differential pumping. The design of the trap cryo-package is shown in detail in Figure \ref{fig:TrapPackage}. The assembly consists of a lid sealed to a ceramic pin-grid array (CPGA) package, on which the ion trap is mounted. The lid is machined out of copper, the side windows are made out of N-BK7 glass, and the imaging window is made out of sapphire. All windows are AR coated and attached to the lid with cryogenic-compatible epoxy (Epotek T7110). The top imaging viewport features a machined trench which serves as a meandering narrow pumping port for evacuating the volume inside the cryo-package at room temperature before the system cools down. The 100-pin CPGA package is modified with a `ringframe' braised to it, which is 1 mm wide and protrudes 2.5 mm from the CPGA surface. This ringframe mates with a matching groove in the lid, designed for a tight fit with indium wires in between. The ringframe serves both as an alignment mechanism for the lid and as mechanical reinforcement for their mating. 
    
    The indium wires increase the thermal conduction between the lid and the CPGA package, and serve as a breakable mechanical bond between the two pieces.
    
    The top of the lid features eight tapped holes for affixing the lid firmly to the cold finger sample mount (Figure \ref{fig:TrapPackage}(a). This top surface is the main thermal and mechanical anchor for the assembly on the sample mount. An activated carbon getter material is placed on the lid inside the UHV volume. Upon cooldown, the lid is cooled most efficiently, and the cryopumping by the getter achieves a UHV environment within the sealed enclosure. 
    
    Some important internal features of the lid can be seen in Figure \ref{fig:TrapPackage}(b). The first is an oblong conical ground shield feature which is machined beneath the imaging window to allow the full numerical aperture of the imaging optics to be available while maximally shielding the potential impact of charge buildup at the exposed dielectric window from influencing the ions. There is also an internal holder to house a ytterbium ablation target. This ablation holder has optical access from a window directly across from the target, such that the path of atomic flux is orthogonal to the CW laser beam path for photoionization lasers to enable isotope-selective loading.\cite{Vrijsen:19}
    
    The trap (Sandia HOA 2.0 \cite{maunz2016high}) is mounted on an interposer, placed on a ceramic spacer to bring the trap surface in line with the windows of the lid. The interposer also contains trench capacitors (TCs), to filter the RF signal on the DC electrodes. The HOA 2.0 trap we used in the experiment had an additional layer of gold (0.5 \(\mu\)m) deposited on the surface of the electrodes. 
    
	The trap cryo-package is assembled in a clean room to minimize dust contamination and is installed into the printed circuit board (PCB) by pressing it into the 100-pin socket. This PCB and trap cryo-package assembly is installed into the 5K sample mount in the cryostat chamber. 
	
\subsubsection{Cryostat Chamber Interior Design}

	\begin{figure}[ht]
        \centering
        \includegraphics[width=\columnwidth]{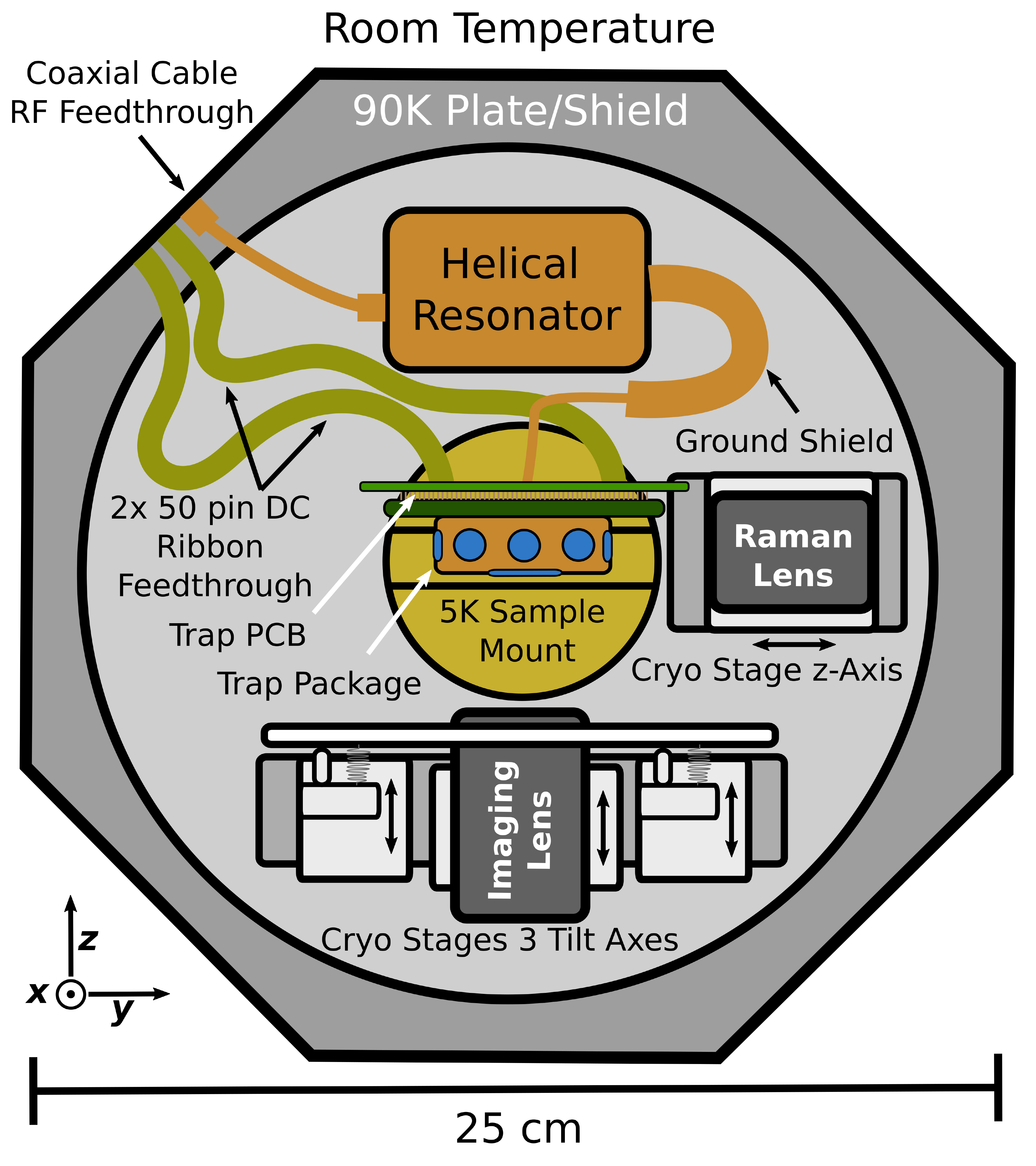}
        \caption{Schematic representation of the cryostat interior, showing the helical resonator, the two interior cryo lenses, and the electrical feedthroughs. The trap axis is along the x-axis, and all lasers are brought to the ion in the x-y plane. }
        \label{fig:CryoInside}
    \end{figure}    
    
    Figure \ref{fig:CryoInside} shows a schematic diagram of the interior elements included in the sample chamber of the cryostat. The trap cryo-package assembly and the PCB are installed into the 5K sample mount with a small amount of thermal grease placed at the interface. The copper assembly lid makes the mechanical and thermal contact with the sample mount, and additional screws are used to support the PCB for strain relief. The PCB has a pair of thermally lagged DC cables and the RF signal pin.
    
    Inside the sample chamber, we install three custom experimental elements on the 90K stage. First is the RF helical resonator, with a coaxial cable input and a simple thick copper wire output. It is mounted close to the trap to avoid long cable between the resonator and the trap.
    
    Also mounted on the 90K stage are two lenses specially designed for operation at cryogenic temperatures. The first is a 0.6 NA ion imaging lens (Photon Gear) with a 150 \(\mu\)m field of view. This is mounted inside the cryostat to achieve a short working distance to the ions. In order to align this lens we designed a specialized alignment stage. It has two micrometer-adjusted X and Y positioners which are adjusted at room temperature. The stage also includes a custom kinematic-mount, with three cryogenic precision actuators (Attocube Inc.) that enables the adjustment of the tip, tilt and z-translation of the lens, which are critical knobs to minimize imaging aberrations. The imaging lens is designed to be an infinite conjugate lens that allows us to use any off-the-shelf lens outside the cryostat to form an image with desired magnification. The infinite conjugate design also reduces aberrations that would arise due to traveling through the 90K and room temperature view-ports. 
    
    The second lens is a Raman projection lens (Photon Gear) for the final tight focusing of individual addressing beams to the ion chain. We can achieve a ~\(1.5 \mu\)m focused beam waist at the ion with this lens. This lens is placed manually at room temperature, with an \textit{in-situ} translation stage for the focus control. 
    
\subsubsection{RF Resonator}

\begin{figure}[ht]
        \centering
        \includegraphics[width=\columnwidth]{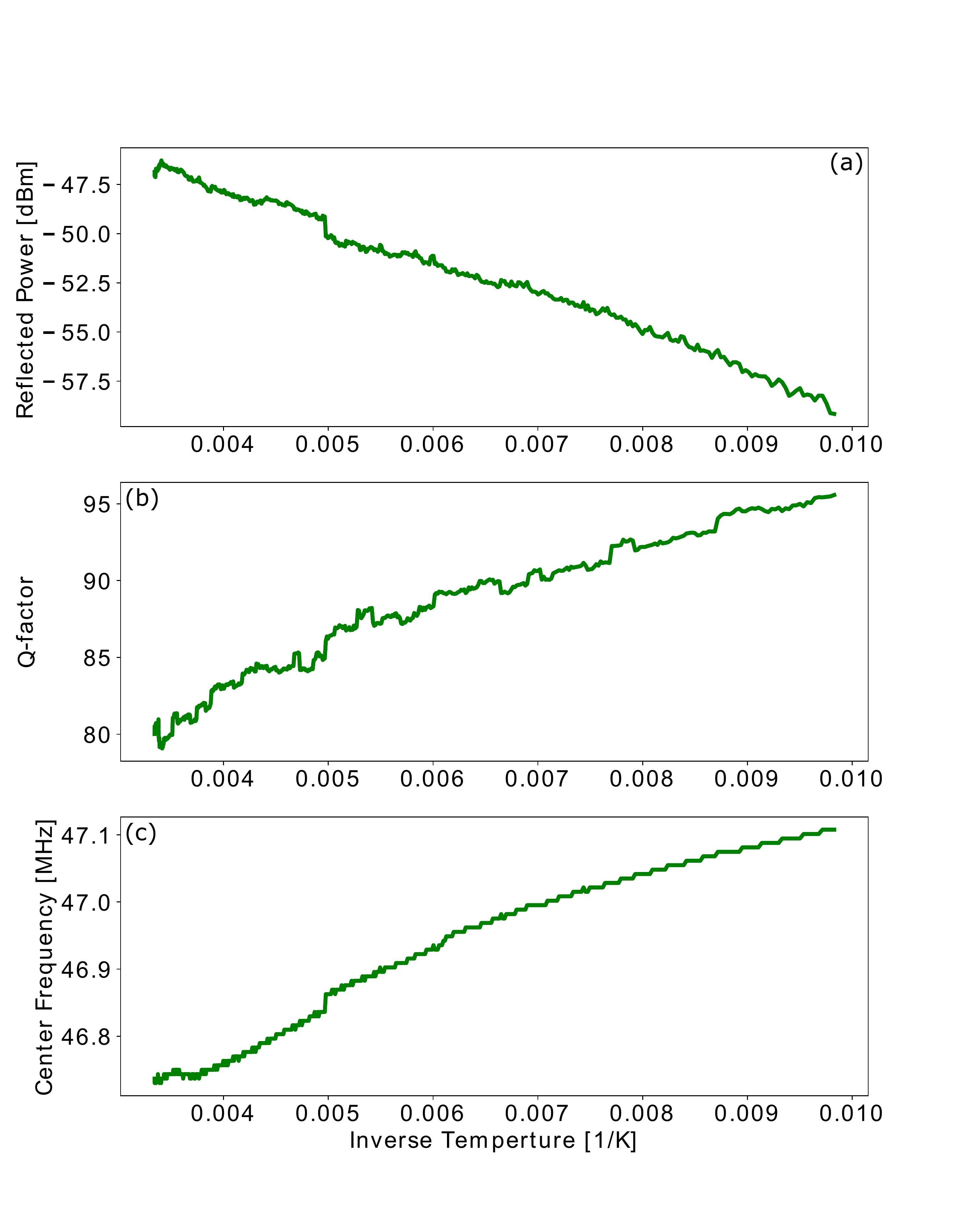}
        \caption{The behavior of the helical resonator as a function of temperature during the cryostat cooldown procedure, which typically takes about 8 hours. The data is plotted in inverse temperature to guide the eye in the direction of time, and we observe the cooling as it passes between 300K and 100K as measured at the sample thermometer. (a) The reflected power goes down as the cryostat cools because we intentionally detune the matching micrometer from optimal so that thermal changes will bring it closer to the matching condition. (b) The quality factor increases during the cooling cycle, reaching ~120 at the base temperature (not shown). (c) The center frequency of the resonator increases as the temperature decreases.}
        \label{fig:RFCool}
    \end{figure}

   The helical resonator is also mounted on the 90K stage within the cryostat, and has a compact design with an overall length of 6.9 cm, a coaxial cable input, and thick copper wire output. The copper wire has a grounded shield made of copper piping bridging most of the length between the helical resonator and the trap PCB RF pin. Figure \ref{fig:RFCool} shows a measurement of the behavior of relevant RF resonance parameters during the cooldown process from room temperature to 100K. It is plotted in inverse temperature to guide the eye in the direction of time during the cooling process. As the system cools, the  Q-factor increases, the reflected power decreases, and the center frequency increases. The increase in Q is expected due to reduced resistive losses in the resonator. The reflected power goes down because the impedance matching condition is intentionally detuned to compensate for the changes during the cooling process. After the cooling process is complete and the system settles at its base temperature, we measure a Q-factor of 120.
    
\subsection{System Characterization}
The optomechanical system is designed to achieve mechanical and thermal stability sufficient to implement high fidelity quantum logic gates. This requires that the optical wavefronts of the intersecting Raman beams be stable at the ion location over the duration of entangling gates that utilize the motional degrees of freedom of the ions in the chain \cite{lee2005phase, inlek2014quantum, wu2018noise}. This level of stability is challenging to achieve using closed-cycle cryostats due to their inherent mechanical vibrations. In this section we present the characterization of our system stability.
    
\subsubsection{Thermal Performance}
    The cryostat used in this work incorporates a Sumitomo RDK-101D cryocooler. Sample space is rigidly attached to the base, and a unique cold finger design provides the thermal link between the cryocooler and the sample space while isolating the transfer of vibrations (details are Montana Instruments proprietary information). 
    
    In order to operate the trap cryo-package below 10K to benefit from cryo-pumping by the carbon getter, the heat load to the 5K platform should be kept below 1W. The main heat load in our setup comes from the DC and RF wires, and the applied RF signal for trapping. Since the wires originate from room temperature, special care should be taken to thermally lag these wires at the 90K stage to minimize heat transfer to the 5K sample mount. A summary of operating temperatures for different configurations is given in Table \ref{table:Temperatures}. The measurement shows that the thermal dissipation from the RF signal is a significant contributor to the overall heat load. 
    
    \begin{table}[ht]
    \centering
    \begin{tabular}{||c | c c ||} 
     \hline
     \multicolumn{3}{||c||}{Temperature Readings (Kelvin)} \\
     \hline
     Configuration & Platform & Sample \\ [0.5ex] 
     \hline\hline
     Base temperatures & 4.99 & 5.70 \\ 
     DC cables connected & 6.05 & 6.56 \\
     DC and RF cables connected & 6.20 & 6.75 \\
     RF on (gold on fused silica trap) & 6.51 & 7.05 \\
     RF on (HOA 2 Series) & 8.16 & 8.79 \\  [1ex] 
     \hline
    \end{tabular}
    \caption{A summary of the temperature readings on the 5K platform and in the sample mount for different experimental configurations. After measuring the base temperature without any heat load to the setup, we investigate the effects of extra heat loads in subsequent measurements. We require 96 DC wires for operating with surface traps in our setup. To minimize heat transfer to 5K stage, these wires are made from low thermal conductance metal. Additionally, we need to make a RF connection between helical resonator and the trap PCB, which provides a heat transfer path from 90K to 5K stage. These two connections increases our operating temperature by about 1K. In order to trap Yb ions and work at 2-3 MHz radial motional frequencies, we need to apply RF with 200-300V amplitude.}
    \label{table:Temperatures}
    \end{table}

\subsubsection{Vibrational Performance}
    
\begin{figure}[ht]
\begin{center}
\includegraphics[width=7cm]{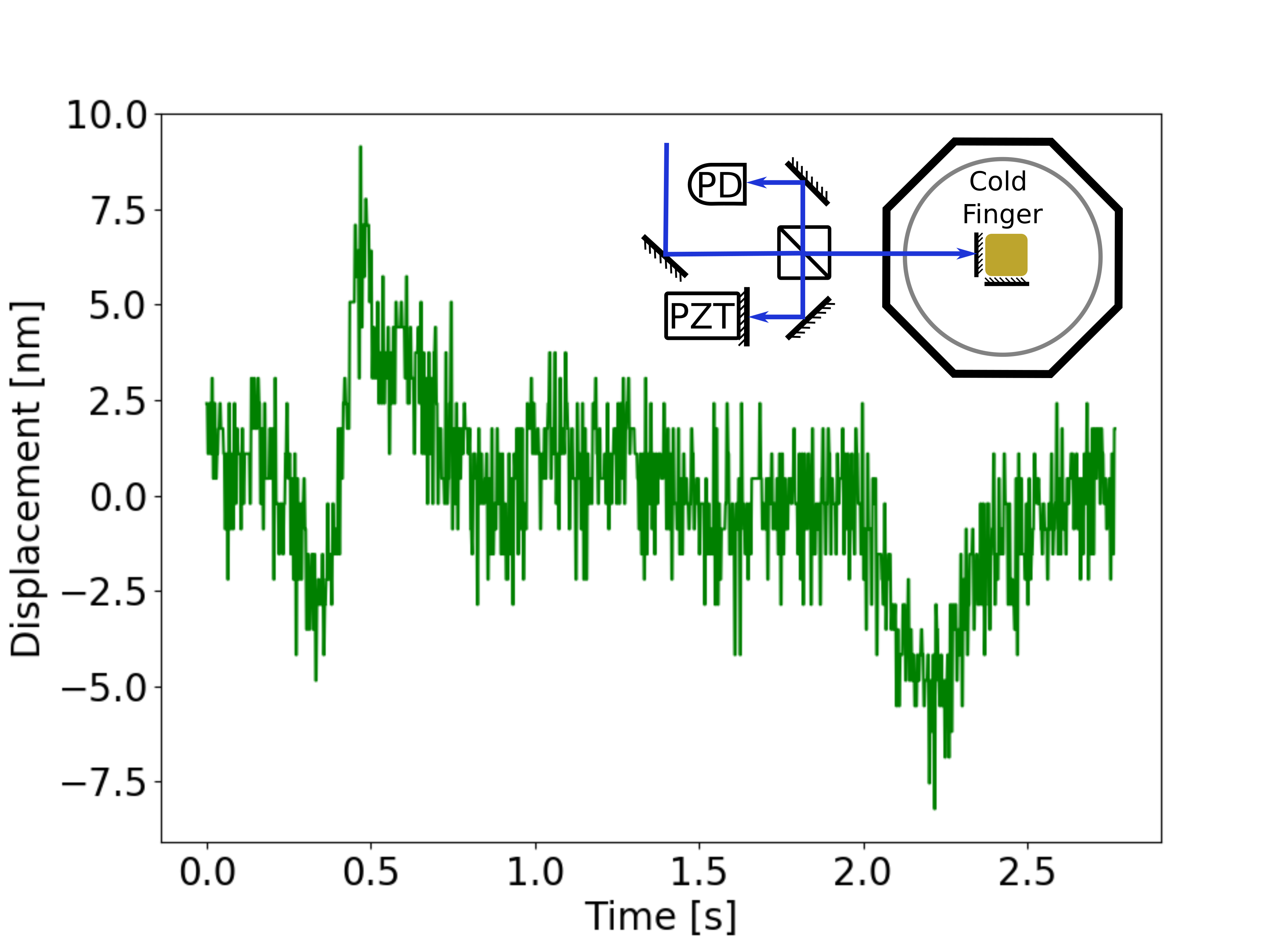}
\caption{Results of the vibration measurement. A Michelson interferometer is used to measure the vibration between optical modules and the cold finger. In our system we experience approximately 17 nm peak-to-peak vibration with an RMS deviation of 2.4 nm. }
\label{fig:vibrationresults}
\end{center}
\end{figure}

    Mechanical vibrations between the trap and Raman lasers cause infidelity in quantum logic gate operations. Conventional closed cycle cryostats feature large mechanical motion of the cold finger sample mounts relative to the lab frame due to the expander motion of the cold head\cite{cryoGV}. We designed an optical interferometer apparatus for characterizing the relative vibrations of our optical modules mounted on the breadboard with respect to the cold sample mount. We constructed an optical module that includes both arms of a Michelson interferometer, where the second arm reflects off of a mirror mounted in place of the ion trap on the cold finger sample mount (inset in Figure \ref{fig:vibrationresults}). We measured the interference signal on a photodiode to invert it into a motional displacement in space. We observe that the displacement due to mechanical vibration is approximately 17 nm peak-to-peak with an RMS deviation of 2.4 nm. This displacement is small compared to the optical wavelength of the Raman beam (355 nm), and should provide stable beams at the ion location to drive high fidelity gates. 

\section{Experiments with Trapped Ions}
	We use \(^{171}\)Yb\(^{+}\) ions as our qubit\cite{ybqubit}. Our system is designed to outperform traditional room-temperature trapped ion experimental setups in providing stable operational environment to achieve higher fidelity quantum computing operations. We characterized the vacuum quality, motional heating rate, and Raman carrier coherence times of ions trapped inside the system, all critical performance metrics needed for high fidelity operations. 
	
\subsection{Vacuum Characterization with Zig-zag Chains}
	The core trap cryo-package used in our system is too compact to permit the integration of an ion gauge or cold-cathode gauge to monitor local pressure. This motivates the development of methods to use the behavior of ions to calculate pressure. One method is to utilize a double well quartic potential of known well-height and then capturing the hopping rate of the ion\cite{aikyo2020vacuum}. A double well potential measurement was problematic in our system due the low potential barriers necessary to probe the low energy collisions at cryogenic temperatures. The noise on the digital-to-analog converters (DACs) was comparable to this well height and it would create a drifting well balance. In order to estimate the vacuum level, we employed a different method utilizing the zig-zag shaped chain of ions.\cite{pagano2018cryogenic} 
	
	A long linear chain of ions can be trapped in a surface trap and the axial confinement potential can be ramped up until a threshold is exceeded. Upon crossing this threshold the chain buckles and collapses into a zig-zag shape (zig-zag phase transition)\cite{schiffer1993phase}. Due to the slight asymmetry in the two radial modes of the trap, there are two energetically degenerate, stable ground states that can arise in a zig-zag shape, whose ion positions are the mirror images of each other about the x-axis (trap axis). We capture images of 7 ion zig-zag chains and count the rate \(\gamma_{zz}\) at which they flip between the two stable configurations. \(\gamma_{zz}\) is related to the molecular collision rate \(\gamma\) by \(\gamma_{zz} = p_{flip}\gamma\), where \(p_{flip}\) is the probability of a zig-zag flip event given a collision with the background molecules.
	
	If \(p_{flip}\) can be estimated as a function of temperature, then this information can be used to translate the observed flipping events between the two zig-zag configurations into molecular collision events, and from these collision events a Langevin model can be used to estimate the pressure inside the cryo-package assembly. The energy barrier between the two zig-zag configurations determines the behavior of \(p_{flip}(T)\). The small energy barrier between the two degenerate modes in the potential can be estimated from geometric considerations.
	
	We estimate the energy barrier between the two zig-zag configurations by considering the energy it would take to rotate a zig-zag chain by \(180^{\circ}\) about the x-axis between the two mode shapes, as this represents a path that the chain could take from one state to the other. We estimated this energy barrier by first calculating the zig-zag structure for seven ions in a 3D harmonic potential. The strength of the confining potential in each axis is measured via RF tickle spectroscopy\cite{noek2013high} of the common modes of motion of a zig-zag chain under the same trapping potential used for the experiments. A simplex optimization algorithm is then used to relax ions into a zig-zag shape in the potential. Once the lowest energy state is found we rotate the chain about the x-axis by \(90^{\circ}\) which will be its highest energy rotational position about this axis. We use our potential energy calculation tool to then calculate the energy difference between the \(0^{\circ}\) and \(90^{\circ}\) position to estimate the `barrier height' between the two modes. In the type of zig-zag potential we used for this experiment, we calculated this energy difference to be approximately 0.02 meV.
	
\begin{figure}[ht]
\begin{center}
\includegraphics[width=1.0\columnwidth]{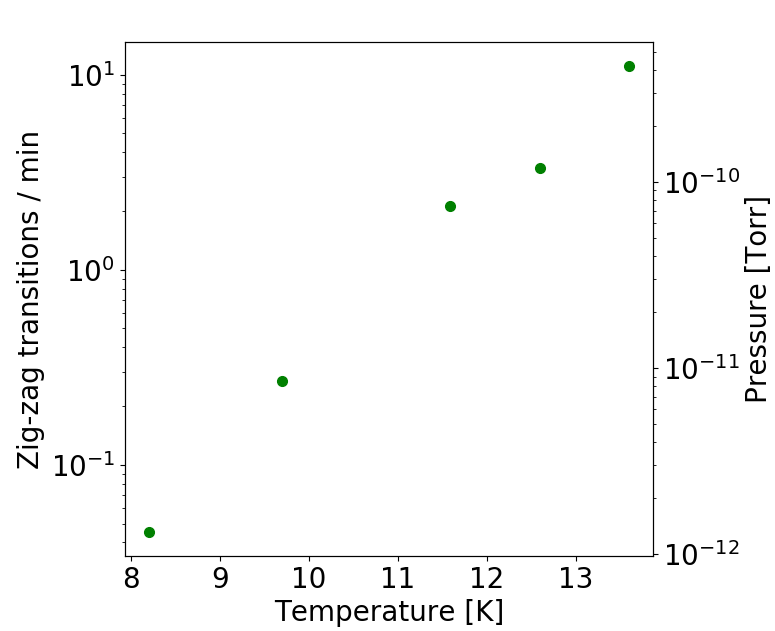}
\caption{On the left axis we see the transition rate between the two zig-zag configurations measured in the experiment as a function of the sample temperature. On the right axis we see the data re-scaled to pressure using Eq. (\ref{eq:Langevin}). }
\label{fig:ZZplot}
\end{center}
\end{figure}

 The effective kinetic energy of the colliding background molecule (predominately H\(_2\)) is scaled by a factor of $\sim$50 from the value calculated\cite{pagano2018cryogenic} using the ion and H\(_2\) masses to account for the fact that many collisions are glancing and don't transfer all of their energy to the ion chain. This is factored into the model as a larger effective threshold kinetic energy for a comparable molecular distribution at the estimated temperature. Using this effective threshold energy \(E_{\mathrm{th,eff}}\), we estimate the population of background molecules that could lead to the transition between the two zig-zag configurations as a function of temperature. Ultimately this analysis leads to a scaling behavior given by
	\begin{eqnarray}
	\label{eq:ZZ}
	P_{zz}(T,E_{\mathrm{th}}) & = & \frac{\int^{\infty}_{ E_{\mathrm{th,eff}}} (\frac{m_{H_2}}{2\pi k_BT})^{3/2}4\pi (\frac{2E_k}{m_{H_2}}) \exp (\frac{-E_k}{k_B T}) dE_k}{\int^{\infty}_{0} (\frac{m_{H_2}}{2\pi k_BT})^{3/2}4\pi (\frac{2E_k}{m_{H_2}}) \exp (\frac{-E_k}{k_B T}) dE_k } \nonumber \\
	& = & \frac{\int^{\infty}_{E_{\mathrm{th,eff}}}E_k\exp (\frac{-E_k}{k_B T}) dE_k}{\int^{\infty}_{0} E_k \exp (\frac{-E_k}{k_B T}) dE_k }, 
    \end{eqnarray}
	where \(E_k\) is the kinetic energy of the background molecule. Using this probability of transition between the zig-zag configurations given a collision, we can invert the zig-zag transition data to estimate the collision rate \(\gamma\) as a function of temperature. Finally, we use \(\gamma\) and a classical Langevin model to calculate the pressure P as
	\begin{equation}
	\label{eq:Langevin}
	     P = \frac{\gamma_{zz}(T)k_B T}{p_{flip}(T) e}\sqrt{\frac{\mu \epsilon _0}{\alpha_{H_2} \pi}},
	\end{equation}
	where \(\alpha_{H_2}\) is the static polarizability of H\(_2\) molecules,\cite{gough1996analysis} \(e\) is the electron charge, \(\mu\) is the reduced mass of the H\(_2\) and \(^{171}\)Yb\(^+\) system, \(\epsilon _0\) is the permittivity of free space. Figure \ref{fig:ZZplot} shows the results of this analysis. On the left-hand side we show the transition rate between the two zig-zag configurations observed (events per minute). On the right-hand side we plot this transition rate re-scaled by Equation \ref{eq:Langevin} to estimate the pressure inside the UHV assembly. We see that the pressure scales with temperature exponentially. At cryo-UHV conditions, the error of pressure estimation is large because the chain transition rate is extremely low (a single event over several hours). Elevating the system temperature and observing the increasing prevalence of zig-zag flipping events allows one to trace out an estimated pressure curve. This pressure curve is most useful to characterize the onset of the regime in which transitions between the zig-zag configurations are prevalent. This exponential scaling with temperature arises because of the exponentially activated desorption of H\(_2\) molecules from the activated carbon getter. 
	
\subsection{Motional Heating Rate}
The motional heating rate of ions in the trap is of paramount importance for two-qubit gate fidelity as most entangling gates utilize the motional degrees of freedom to mediate the qubit interaction. The gate fidelity and circuit depth attainable in our system will also be effected by the heating rate \cite{wang2020high}. The measurement scheme is similar to that used in Mount et al. \cite{turchette2000heating,mount2013single}. We measure the heating rates of the radial modes with a ~45\(^\circ\) trap rotation so that the Raman beams can interact with both radial modes. The ion motion is first cooled by Doppler cooling so that average motional excitation  is $\leq$5 quanta, followed by 30 rounds of resolved sideband cooling\cite{leibfried2003quantum} to reach an average motional excitation of $\sim$0.09 quanta in the radial modes. After waiting for a duration, we drive both the red and blue sideband transitions on resonance. 
We perform a fit to the experimental data with a simulated Hamiltonian of the blue side-band rotations with heating decoherence effects. We include an exponential decay envelope to simulate the motional dephasing effects. We use the fitted value of \(\bar{n}\) of these sideband rotations to determine the average motional excitation (or temperature) at any given wait time. 

\begin{figure}[ht]
\begin{center}
\includegraphics[width=7cm]{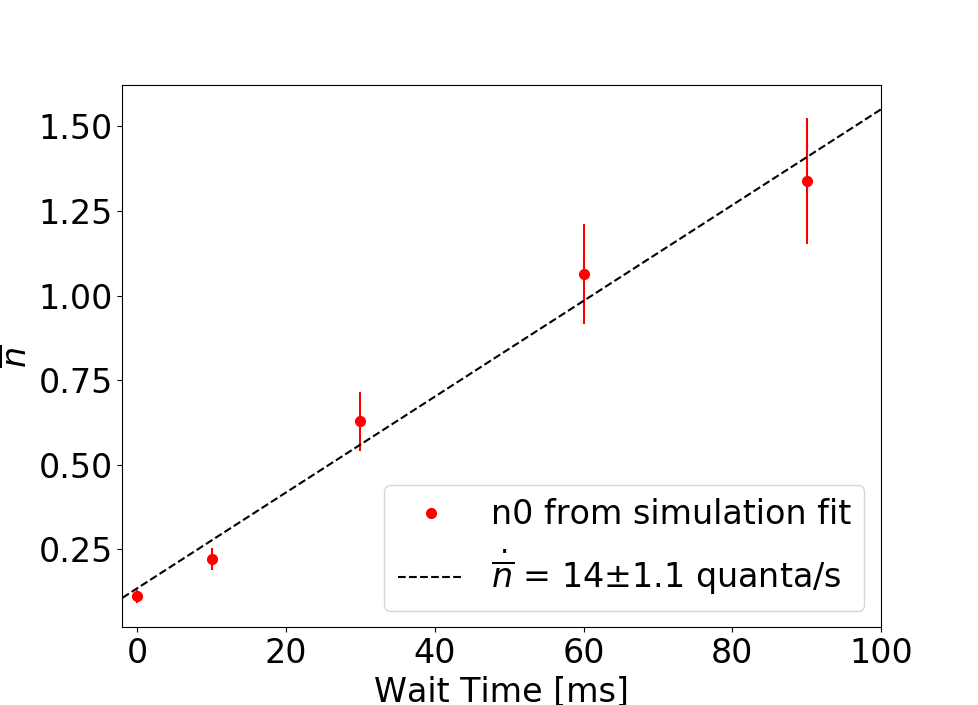}
\caption{Heating rate measurement at a 2.23 MHz motional frequency. We measure a heating rate of 14\(\pm\)1.1 quanta/s with a maximum wait time of 90 ms}
\label{fig:HeatingRate}
\end{center}
\end{figure}

We repeat this experiment for a variety of wait times, during which the surface noise influences the ion and causes heating. We scan the wait time to trace out the average motional excitation versus wait time and fit the results to establish a heating rate. The error of this heating is established from the covariance of the linear fit (Figure \ref{fig:HeatingRate}). We measured the heating rate at 15 different trap mode frequencies between 1.2 and 2.3 MHz, and found no apparent frequency trend. The lack of frequency trend could indicate that we are still limited by technical noise. The lowest measured heating rate was 5.0\(\pm\)0.8 quanta/s at 1.82 MHz trap mode frequency, and the largest measured heating rate was 25\(\pm\)4.3 quanta per second at 2.04 MHz. The mean heating rate of these measurements was 13 quanta/s with a standard deviation of 5.1 quanta/s. A representative plot at operational trap frequencies for two-qubit gates on the radial modes is presented in Figure \ref{fig:HeatingRate}, where we measure a heating rate of 14\(\pm\)1.1 quanta/s. Improvements in the Raman intensity stability could reduce the measurement noise enough to detect a scaling trend with confidence; however for this data the trend seems flat within the range of investigation.

In order to compare this to similar cryogenic systems, we need to normalize for the ion species and the trap frequency. Similar to Hite et al.~\cite{ArgonIonHeating} we can calculate the electric field noise spectral density \(S_{E}(\omega)\) as
\begin{equation}
    \label{eq:EFNC}
    S_{E}(\omega) = \frac{4m\hbar \omega}{q^2}\dot{\bar{n}},
\end{equation}
where $m$ is the ion mass and $q$ is the elementary charge.
When we multiply this by the motional frequency to obtain \(\omega S_{E}(\omega)\), we can compare our results to other low temperature surface traps at different trap-ion distances.\cite{ArgonIonHeating} Our noise spectral density \(\omega S_{E}(\omega) =\) 3.2\(\times 10^{-7} V^2/m^2\), which lies an order of magnitude below the average line of preexisting experiments with 70 \(\mu\)m ion-trap distance at cryogenic temperatures. Based on careful considerations of the error contribution to entangling gates, we anticipate that this level of heating rate would not limit the gate fidelity up to 99.99\%\cite{wang2020high} 

\subsection{Coherent Qubit Operations}
The loss of optical coherence in the Raman beams within the duration of the entangling gates limit the fidelity of the gate. \cite{wang2020high} The optical coherence of counter-propagating Raman beams was characterized with our 355 nm laser systems. We performed Ramsey spectroscopy on the carrier Raman rotations to measure the interferometric stability of the Raman beams at the ion location. We first scan the two red sideband motional modes in frequency space to find the resonance condition, then perform 25 rounds of sideband cooling at these radial frequencies. This is followed by a \(\frac{\pi}{2}\) carrier rotation to put the qubit in a coherent superposition state, then at a variable wait time later, the second \(\frac{\pi}{2}\) rotation is performed with a shifted phase. This phase is scanned over 11 different values to trace out the parity curves to measure the loss of coherence by extracting the contrast in the Ramsey interference fringes. If the optical phase of the Raman beams seen by the ion between the two \(\frac{\pi}{2}\) pulses fluctuate during the wait time, the contrast of the parity curve will degrade.   

\begin{figure}[ht]
\begin{center}
\includegraphics[width=1\columnwidth]{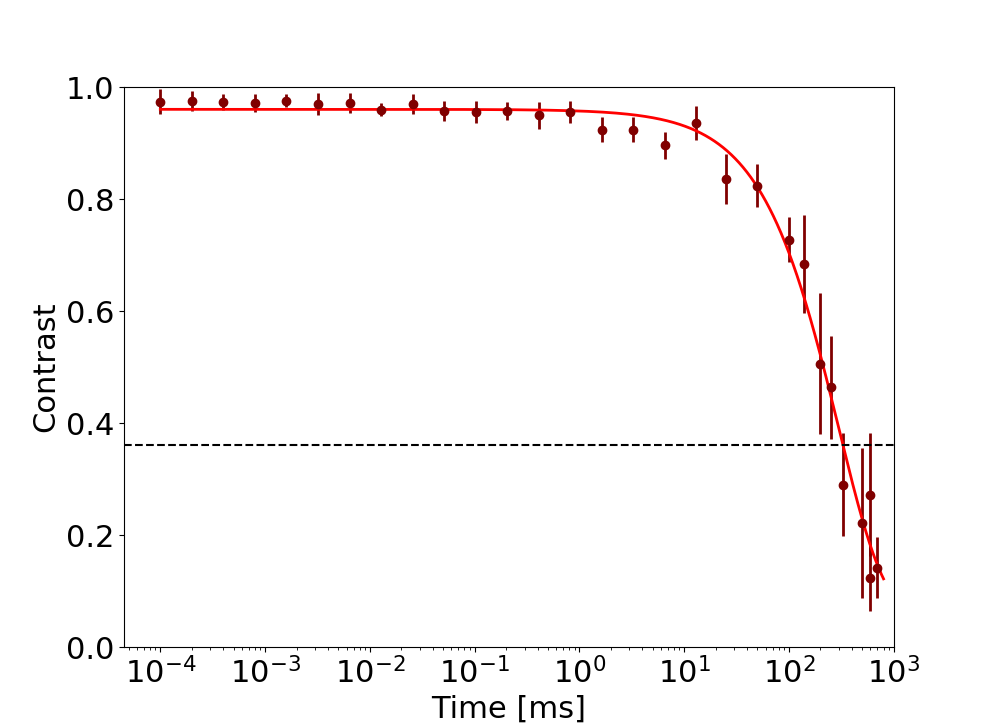}
\caption{Counter propagating carrier transitions used to perform a Ramsey experiment. The coherence time obtained from this measurement is 330 ms.}
\label{fig:Ramsey}
\end{center}
\end{figure}

This experiment was repeated and the wait time was scanned out to 1 second. The results of this experiment (without using spin echo) are shown in Figure \ref{fig:Ramsey}. Fitting this fringe contrast to an exponential decay, we estimate the \(\frac{1}{e}\) decay time to obtain the optical coherence time of 330 ms. We do not expect the optical coherence of our system to be a limiting factor for the entangling gate fidelity\cite{wang2020high}.

\section{Conclusions}

We have shown the design of an ion trap system based on a compact cryogenic package. An HOA 2.0 trap with a 0.5 \(\mu\)m thick gold re-coating was installed in this system, and supports sufficiently high RF amplitude to achieve radial mode frequencies of 2-3 MHz.  The system includes a high NA lens and helical RF resonator inside the cryostat on the 90K plate. This system is tightly integrated into a surrounding optomechanical structure in which all free-space optical blocks are attached and precision aligned.

Differential pumping with cryo pumping creates state-of-the-art UHV levels necessary to hold long ion chains for many hours at a time. We used zig-zag chains of ions to map out an estimated pressure curve as a function of sample temperature.  

The motional heating rate in the HOA 2.0 trap was measured the mean heating rate over a range of motional frequencies was found to be 13 quanta/s, when normalized for mode frequency and ion species, gives the electric field noise spectral density \(\omega S_{E}(\omega) =\) 3.2\(\times 10^{-7} V^2/m^2\). This value is significantly better than other examples\cite{ArgonIonHeating}, and should not pose any limitation in achieving high-fidelity entangling gates. 

We have demonstrated a 330 ms optical coherence time for counter-propagating carrier rotations, which demonstrates the optical stability of the setup despite operating in a closed-cycle cryostat.

\section{Acknowledgements}
This work was primarily supported by the Office of Director of National Intelligence - Intelligence Advanced Research Projects Activity through ARO contract W911NF-16-1-0082 (experimental implementation), and National Science Foundation STAQ projects PHY-1818914 (vibration characterization).

\bibliography{CryoSystem.bib}

\end{document}